\documentclass[letterpaper, 10pt]{article}
\usepackage[utf8]{inputenc}
\usepackage[top=1in, bottom=1in, left=1in, right=1in]{geometry}
\usepackage{float}
\usepackage{color}
\usepackage{tikz}

\usepackage{titling}
\RequirePackage[explicit]{titlesec}
\titlespacing*{\section}{0pt}{0.5em}{0.3pt}
\titlespacing*{\subsection}{0pt}{0.35em}{0pt}
\titlespacing*{\subsubsection}{0pt}{0.25em}{0pt}

\usepackage{caption}
\RequirePackage{fancyhdr}
\usepackage{lastpage}
\usepackage{empheq}

\usepackage{mathpazo}
\usepackage{xcolor}
\definecolor{GT}{RGB}{0, 0, 0}
\definecolor{Conv}{RGB}{247,37,133}
\definecolor{MC}{RGB}{114,9,183}
\definecolor{MVN}{RGB}{67,97,238}
\definecolor{FE}{RGB}{76,201,240}

\usepackage[
  colorlinks=true,
  linkcolor=blue,
  anchorcolor=blue,
  citecolor=blue,
  filecolor=blue,
  menucolor=blue,
  runcolor=blue,
  urlcolor=blue]{hyperref}
\usepackage{amsmath, amsthm, amssymb, amsfonts}
\usepackage{upgreek}
\usepackage{dsfont}

\usepackage{enumerate}
\usepackage{enumitem}

\usepackage{graphicx}
\usepackage{subcaption}

\usepackage{multirow}

\usepackage[numbers, sort&compress]{natbib}

\title{Flow Reconstruction and Particle Characterization from Inertial Lagrangian Tracks}

\author{
  Ke Zhou and
  Samuel J. Grauer\thanks{Corresponding author: \href{mailto:sgrauer@psu.edu}{sgrauer@psu.edu}}\vspace*{.15em}\\
  {\small Department of Mechanical Engineering, Pennsylvania State University}}
\date{}

\begin{document}

\maketitle
\setcounter{footnote}{0}
\vspace*{-2em}

\begin{abstract}
\noindent This text describes a method to simultaneously reconstruct flow states and determine particle properties from Lagrangian particle tracking (LPT) data. LPT is a popular measurement strategy for fluids in which particles in a flow are illuminated, imaged (typically with multiple cameras), localized in 3D, and then tracked across a series of frames. The resultant ``tracks'' are spatially sparse, and a reconstruction algorithm is commonly employed to determine dense Eulerian velocity and pressure fields that are consistent with the data as well as the equations governing fluid dynamics. Existing LPT reconstruction algorithms presume that the particles perfectly follow the flow, but this assumption breaks down for inertial particles, which can exhibit lag or ballistic motion and may impart significant momentum to the surrounding fluid. We report an LPT reconstruction strategy that incorporates the transport physics of both the carrier fluid and particle phases, which may be parameterized to account for unknown particle properties like size and density. Our method enables the reconstruction of unsteady flow states and determination of particle properties from LPT data and the coupled governing equations for both phases. We use a neural solver to represent flow states and data-constrained polynomials to represent the tracks (though we note that our technique is compatible with a variety of solvers). Numerical tests are performed to demonstrate the reconstruction of forced isotropic turbulence and a cone--cylinder shock structure from inertial tracks that exhibit significant lag, streamline crossing, and preferential sampling.\par\vspace{.5em}

\end{abstract}
\vspace*{2em}

\section{Introduction}
\label{sec:intro}
Lagrangian particle tracking (LPT) is a powerful tool for making 3D measurements of complex flows involving turbulence \cite{Bross2023, Zhou2023}, flow separation \cite{Schroder2020, Schroder2015}, variable density mixing \cite{Weiss2023, Tan2023}, and more \cite{Schroder2023}. Compared to tomographic particle image velocimetry (PIV), LPT offers direct measurements of particle positions, nearly-ghost-free particle detection \cite{Schanz2016}, and accurate pressure estimation \cite{vanGent2017}. The general procedure for LPT goes as follows. Tracer particles are seeded into the flow or arise spontaneously, e.g., droplets, bubbles, snowfall, etc.\footnote{Particles are the ``disperse'' phase and the transport medium is the ``carrier'' phase, collectively called disperse multiphase flow.} These particles are illuminated, usually with a laser; imaged by one or more cameras; and localized to identify their 3D positions in each frame. Next, a tracking algorithm is employed to link individual particles across frames, resulting in a set of Lagrangian particle trajectories, called ``tracks''. These tracks are spatially sparse, so post-processing is often performed to estimate the continuous Eulerian velocity field, termed ``flow reconstruction''. Advanced reconstruction algorithms combine LPT tracks with the equations governing fluid flow to improve the accuracy of velocity field estimates and infer pressure \cite{Jeon2018, Jeon2022, Schanz2018, Ehlers2020}. Existing algorithms assume that the particles are \textit{ideal tracers}, meaning that they passively follow the surrounding flow. However, many applications feature inertial particles that may lag the flow or travel ballistically, and large, heavy, buoyant, or sufficiently numerous particles can even modulate a flow's behavior \cite{Brandt2022}. What's more, particle properties, including their size, density, and morphology, are often unknown in an LPT experiment and cannot be deduced from the image data. We report a new technique that addresses these limitations by simultaneously reconstructing Eulerian flow states and determining particle properties from inertial LPT tracks via the equations that govern disperse multiphase flow.\par

Particle motion in a fluid is a consequence of the particle's inertia and the net force acting on said particle \cite{Mei1996, Maxey1997}. Viscous drag is typically the dominant force for small particles.\footnote{Other hydrodynamic forces, like the added mass, pressure, and Basset history force, are negligible for small, dense particles, which usually holds true for tracers in PIV and LPT experiments \cite{Maxey1997}. See \ref{app:inertial particle velocimetry} for more information.} Drag is induced by viscous forces when there is relative motion between a particle and the surrounding fluid, called ``slip''. The carrier fluid resists slip and speeds or slows the suspended particles to the ambient flow speed. This occurs quickly for small and light particles, resulting in ideal tracer behavior, whereas large and heavy particles respond slowly to slip. In addition, body forces like gravity, buoyancy, and magnetic attraction or repulsion may also induce a slip velocity \cite{Frankel2016, Salibindla2020}. Particle transport can be characterized by a particle relaxation time, $\tau_\mathrm{p}$, defined here as the slip velocity's decay rate following a step change in the flow speed \cite{Melling1997}. The ratio of $\tau_\mathrm{p}$ to the characteristic flow time scale, $\tau_\mathrm{f}$, is the Stokes number, $St = \tau_\mathrm{p}/\tau_\mathrm{f}$. Particles with a vanishing Stokes number are ideal tracers, whereas particles with $St \gtrsim 0.1$ are deemed to have non-negligible inertia and inadequate ``traceability'' for particle-based velocimetry (\textit{when using existing algorithms}) \cite{Raffel2018, Samimy1991}. As mentioned above, high-Stokes-number transport is common in selected LPT and PIV experiments, both in laboratory and natural environments, which can corrupt ensuing reconstructions. Some examples include the unintentional inertial transport of bubble tracers in aerodynamic flows \cite{Wolf2019}; solid particles in shocked flows \cite{Ragni2011}, high-speed boundary layers \cite{Brooks2018}, and jets in crossflow \cite{Nair2023}; snowy atmospheric boundary layers \cite{Li2021}; and sediment transport in waterways \cite{Righetti2004}. \ref{app:inertial particle velocimetry} contains a detailed discussion of inertial particles in the context of LPT and PIV.\par

This text presents a novel reconstruction framework for LPT that accommodates inertial particle tracks, including cases where the particle properties are unknown. Our algorithm seeks flow and particle states as well as particle properties that solve the governing equations for disperse multiphase flow, accounting for the interphase transfer of mass, momentum, and energy. Flow states are parameterized using a coordinate neural network, and the particle tracks are represented with polynomials that incorporate the LPT data as a hard constraint. A trainable set of particle properties -- size, density, etc. -- is included in cases where these properties are unknown. The flow states, particle trajectories, and particle properties are determined by minimizing a set of physics residuals and boundary conditions. We demonstrate the approach using synthetic LPT data from two scenarios: (1)~bidisperse particles suspended in forced isotropic turbulence and (2)~polydisperse particles in supersonic flow over a cone--cylinder body. What follows is a description of the algorithm in Sec.~\ref{sec:method}, numerical results in Sec.~\ref{sec:demos}, and conclusions in Sec.~\ref{sec:conclusions}.\par

\section{Methodology}
\label{sec:method}

\subsection{Neural-implicit particle advection}
\label{sec:method:data assimilation}
\begin{figure}[ht]
\vspace*{-0.5em}
    \centering
    \includegraphics[width=\textwidth]{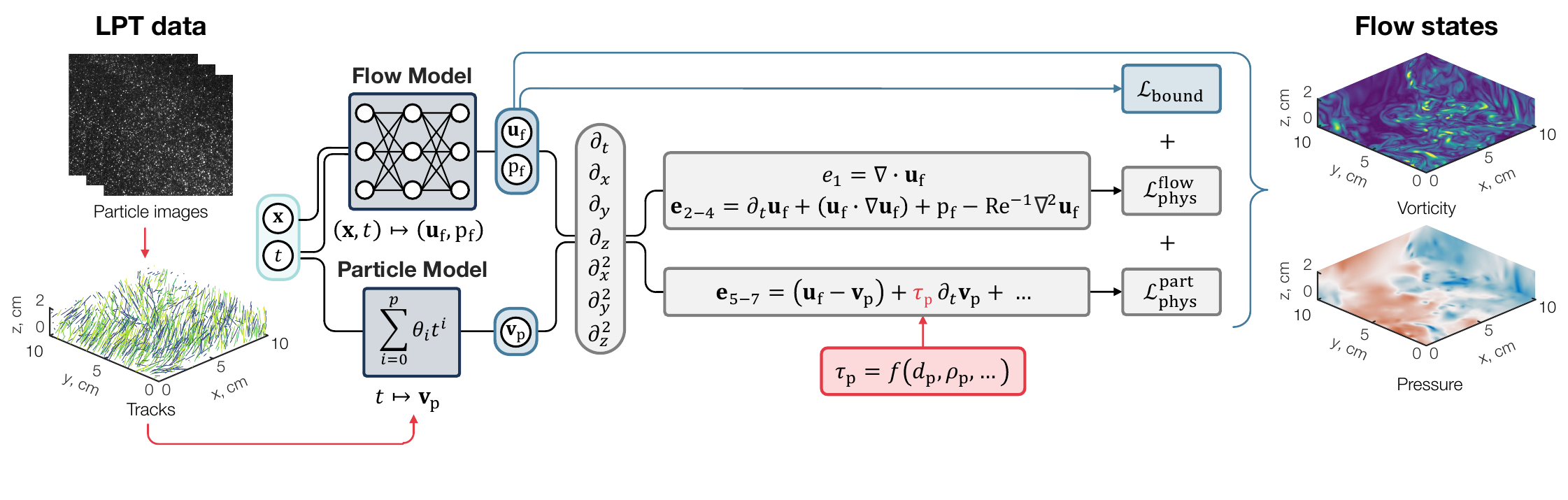}
    \caption{Schematic of our method. Neural flow states represent the carrier phase and constrained polynomials represent particle states, with the LPT data embedded as a hard constraint. Flow and particle states as well as particle characteristics are inferred from the data and governing equations.}
    \label{fig:schematic}
\end{figure}

The proposed method employs neural flow states coupled with constrained polynomials to reconstruct particle-laden flows from LPT data. Figure~\ref{fig:schematic} depicts the overall framework, which is implemented in a machine learning environment. A deep neural network, $\mathcal{F}$, maps space--time coordinates to the flow state at that location,
\begin{equation}
    \label{equ:flow states}
    \mathcal{F}\mathopen{}\left(\boldsymbol\uptheta_\mathrm{f}\right): \left(\mathbf{x}, t\right) \mapsto \left(\mathbf{u}_\mathrm{f}, p_\mathrm{f}, \rho_\mathrm{f}, \dots\right),
\end{equation}
where $\mathbf{x}$ is a spatial location; $t$ is time; $\mathbf{u}_\mathrm{f}$, $p_\mathrm{f}$, and $\rho_\mathrm{f}$ are the carrier phase velocity, pressure, and density fields; and $\boldsymbol\uptheta_\mathrm{f}$ is a vector of network parameters: weights, biases, etc., per Sec.~\ref{sec:method:network}. Note that we consider both 2D and 3D problems, and the flow state can be expanded to include additional variables as needed. The particle advection model uses a polynomial, $\mathcal{P}^k$, for each particle to describe its velocity over time,
\begin{equation}
    \label{equ:particle tracks}
    \mathcal{P}^k\mathopen{}\left(\boldsymbol\uptheta_\mathrm{p}^k\right) : t \mapsto \mathbf{v}_\mathrm{p} \quad\text{for}\quad k \in \{1, 2, \dots, n_\mathrm{p}\},
\end{equation}
where $\mathbf{v}_\mathrm{p}$ is the particle velocity, $k$ is the trajectory index, $n_\mathrm{p}$ is the number of particles, and $\boldsymbol\uptheta_\mathrm{p}^k$ is a vector of polynomial coefficients for the $k$th track. The form of $\mathcal{P}^k$ is elaborated in Sec.~\ref{sec:method:tracks}. In addition to $\boldsymbol\uptheta_\mathrm{f}$ and $\boldsymbol\uptheta_\mathrm{p}^k$, we can specify a trainable vector of properties for each particle, $\boldsymbol\uppsi^k$, which may include their size, density, charge, etc. In a well-characterized LPT experiment, where all the salient particle properties are known in advance or can be determined from the image data, $\boldsymbol\uppsi^k$ is omitted. Trajectory coefficients and properties for all the particles are collected in matrices, denoted $\boldsymbol\Uptheta_\mathrm{p}$ and $\boldsymbol\Uppsi$, respectively.\par

Ideally, the flow states from $\mathcal{F}$ and particle trajectories implied by $\mathcal{P}^k$ should obey the governing equations, satisfy known boundary conditions, and match experimental data. Consistency with the LPT data is guaranteed by hard constraints on $\mathcal{P}^k$ (Sec.~\ref{sec:method:tracks}), and an objective loss is specified to aggregate residuals from the flow and particle physics equations as well as any boundary conditions,
\begin{equation}
    \label{equ:objective loss:total}
    \mathcal{L}_\mathrm{total}\mathopen{}\left(\boldsymbol\uptheta_\mathrm{f}, \boldsymbol\Uptheta_\mathrm{p}, \boldsymbol\Uppsi\right) = \chi_1 \,\mathcal{L}_\mathrm{phys}^\mathrm{flow}\mathopen{}\left(\boldsymbol\uptheta_\mathrm{f}\right) + \chi_2 \,\mathcal{L}_\mathrm{phys}^\mathrm{part}\mathopen{}\left(\boldsymbol\Uptheta_\mathrm{p}, \boldsymbol\Uppsi\right) + \chi_3 \,\mathcal{L}_\mathrm{bound}\mathopen{}\left(\boldsymbol\uptheta_\mathrm{f}\right),
\end{equation}
where $\chi_1$, $\chi_2$, and $\chi_3$ are loss weighting parameters that must be carefully selected. The flow physics loss is
\begin{equation}
    \label{equ:objective loss:flow}
    \mathcal{L}_\mathrm{phys}^\mathrm{flow}\mathopen{}\left(\boldsymbol\uptheta_\mathrm{f}\right) = \frac{d_\mathrm{f}^{-1}}{\left|\mathcal{V} \times \mathcal{T}\right|} \int_\mathcal{T} \iiint_\mathcal{V} \left\lVert\mathbf{e}_\mathrm{f}\mathopen{}\left(\mathbf{x}, t; \boldsymbol\uptheta_\mathrm{f}\right)
    \right\rVert_2^2 \mathrm{d}\mathcal{V} \,\mathrm{d}t.
\end{equation}
In this expression, $\mathcal{V}$ and $\mathcal{T}$ are the spatial and temporal domains, $\mathbf{e}_\mathrm{f}$ is a vector of the governing equation residuals at position $\mathbf{x}$ and time $t$, and $d_\mathrm{f}$ is the the number of equations for the carrier phase, i.e., the dimension of $\mathbf{e}_\mathrm{f}$. Similarly, the particle physics loss is
\begin{equation}
    \label{equ:objective loss:particle}
    \mathcal{L}_\mathrm{phys}^\mathrm{part}\mathopen{}\left(\boldsymbol\Uptheta_\mathrm{p}, \boldsymbol\Uppsi\right) = \frac{d_\mathrm{p}^{-1}}{n_\mathrm{p}} \sum_{k=1}^{n_\mathrm{p}} \left[\frac{1}{\left|\mathcal{T}^k\right|} \int_{\mathcal{T}^k} \left\lVert\mathbf{e}_\mathrm{p}^k\mathopen{}\left(t; \boldsymbol\uptheta_\mathrm{p}^k, \boldsymbol\uppsi^k\right)\right\rVert_2^2 \mathrm{d}t\right],
\end{equation}
where $\mathcal{T}^k \subseteq \mathcal{T}$ is the time segment for the $k$th track, $\mathbf{e}_\mathrm{p  }^k$ is a vector of governing equation residuals for the $k$th particle at time $t$, $\boldsymbol\uptheta_\mathrm{p}^k$ and $\boldsymbol\uppsi^k$ are the $k$th columns of $\boldsymbol\Uptheta_\mathrm{p}$ and $\boldsymbol\Uppsi$, and $d_\mathrm{p}$ is the number of equations for the disperse phase. Boundary losses depend on the specific condition, for instance, a no-slip wall corresponds to the following loss:
\begin{equation}
    \label{equ:objective loss:boundary}
    \mathcal{L}_\mathrm{bound}\mathopen{}\left(\boldsymbol\uptheta_\mathrm{f}\right) = \frac{d_\mathrm{u}^{-1}}{\left|\mathcal{A} \times \mathcal{T}\right|} \int_\mathcal{T} \iint_\mathcal{A} \left\lVert\mathbf{u}_\mathrm{f}\mathopen{}\left(\mathbf{x}, t; \boldsymbol\uptheta_\mathrm{f}\right)
    \right\rVert_2^2 \mathrm{d}\mathcal{A} \,\mathrm{d}t,
\end{equation}
where $d_\mathrm{u}$ is the dimension of $\mathbf{u}_\mathrm{f}$ and $\mathcal{A}$ is the wall surface. We omit boundary losses unless otherwise stated.\par

Partial derivatives of $\mathcal{F}$ and $\mathcal{P}^k$ are obtained by automatic differentiation and used to determine $\mathbf{e}_\mathrm{f}$ and $\mathbf{e}_\mathrm{p}^k$ by evaluating the governing equations. The integrals in Eqs.~\eqref{equ:objective loss:flow}--\eqref{equ:objective loss:boundary} are computed by Monte Carlo sampling. Ultimately, $\mathcal{L}_\mathrm{total}$ is minimized via a backpropagation algorithm, resulting in an approximate solution to the equations that inherently satisfies the LPT data. Governing equations may vary depending on the application; \ref{app:governing eqs} presents the equations used in this work and \ref{app:implementation} includes some important details about our implementation.\par

\subsection{Network architecture}
\label{sec:method:network}
There are numerous methods to represent flow states for LPT data assimilation, which we reviewed in \cite{Zhou2023}. We opt for coordinate neural networks because they carry several advantages in this context, namely: they are easy to implement, they offer significant data compression, and they provide a mesh-free, analytical representation of flow states and their derivatives, which is ideal for optimization. Our networks comprise an input layer, output layer, and series of $n_\mathrm{l}$ hidden layers,
\begin{subequations}
    \label{equ:method:architecture}
    \begin{align}
        \mathcal{F}\mathopen{}\left(\mathbf{z}^0\right) &= \mathbf{W}^{n_\mathrm{l}+1}\left[\mathcal{N}^{n_\mathrm{l}} \circ \mathcal{N}^{n_\mathrm{l}-1} \circ \dots \circ \mathcal{N}^2 \circ \mathcal{G}\mathopen{}\left(\mathbf{z}^0\right)\right]+\mathbf{b}^{n_\mathrm{l}+1},
        \intertext{with}
        \mathbf{z}^l = \mathcal{N}^l\mathopen{}\left(\mathbf{z}^{l-1}\right) &= \sigma\mathopen{}\left(\mathbf{W}^l\mathbf{z}^{l-1} + \mathbf{b}^l\right) \quad\text{for}\quad l \in \{2, 3, \dots, n_\mathrm{l}\}.
    \end{align}
\end{subequations}
The vector $\mathbf{z}^l$ contains the value of neurons in the $l$th layer, $\mathbf{W}^l$ and $\mathbf{b}^l$ are the weight matrix and bias vector for the $l$th layer, and $\sigma$ is a nonlinear activation function that is applied to each element of the argument. The vector $\boldsymbol\uptheta_\mathrm{f}$ contains all the trainable weights and biases in $\mathcal{F}$. We use swish activation functions,
\begin{equation}
    \label{equ:method:swish}
    \sigma(z) = \frac{z \,\exp(z)}{1 + \exp(z)},
\end{equation}
which have been shown to improve the stability of gradient flow in training compared to hyperbolic tangent functions, among others \cite{Molnar2022}. Moreover, to mitigate the spectral bias of gradient-descent-type training \cite{Wang2021}, we replace the $\mathcal{N}^1$ layer with a Fourier encoding \cite{Tancik2020},
\begin{equation}
    \label{equ:method:FE}
    \mathbf{z}^1 = \mathcal{G}\mathopen{}\left(\mathbf{z}^0\right) = \left[\sin\mathopen{}\left(2\pi \boldsymbol\upomega_1 \cdot \mathbf{z}^0\right), \,\cos\mathopen{}\left(2\pi \boldsymbol\upomega_1 \cdot \mathbf{z}^0\right), \dots, \,\sin\mathopen{}\left(2\pi \boldsymbol\upomega_w \cdot \mathbf{z}^0\right), \,\cos\mathopen{}\left(2\pi \boldsymbol\upomega_w \cdot \mathbf{z}^0\right)\right]^\top.
\end{equation}
In this layer, $w$ is the number of Fourier features and $\boldsymbol\upomega$ is a vector of random frequencies -- fixed before training -- with a unique frequency for each element of $\mathbf{z}^0$.\par

\subsection{Data-constrained tracks}
\label{sec:method:tracks}
Particle tracks contain a sequence of 3D positions: $\mathbf{x}_j^k$ for $j \in \{0, 1, \dots, n_k - 1\}$, where $\mathbf{x}_j^k$ indicates the $k$th particle's location at time $t_j$ and $n_k$ is the total number of positions in that track. By definition, the tracked locations must satisfy an advection equation,
\begin{equation}
    \label{equ:particle advection}
    \frac{\mathrm{d}\mathbf{x}_\mathrm{p}}{\mathrm{d}t} = \mathbf{v}_\mathrm{p} \;\Longleftrightarrow\; \mathbf{x}_j^k = \int_{t_{j-1}}^{t_j} \mathbf{v}_\mathrm{p}\mathopen{}\left(t\right) \mathrm{d}t + \mathbf{x}_{j-1}^k,
\end{equation}
where $\mathbf{x}_\mathrm{p}$ is the position of a particle moving with velocity $\mathbf{v}_\mathrm{p}$. Per equation Eq.~\eqref{equ:particle tracks}, we represent the trajectory of each particle with a polynomial, $\mathcal{P}^k$, which describe the particle's velocity as a function of time.\footnote{$\mathcal{P}^k$ fully determines a particle's trajectory in state space -- position, velocity, acceleration -- relative to $\mathbf{x}_0^k$.} We specify $\mathcal{P}^k$ using a vector of \textit{unconstrained} free parameters, $\boldsymbol{\uptheta}_\mathrm{p}^k$, and a formulation that embeds Eq.~\eqref{equ:particle advection} as a \textit{hard constraint}. The resulting particle states are easy to optimize using backpropagation. For ease of notation, we present a single component of velocity, $v_\mathrm{p}$, and drop the particle index, $k$, for the rest of this section. The extension to 2D and 3D polynomials is trivial.\par

Following the theory of functional connections \cite{Leake2022}, which is a general framework for converting a constrained optimzation problem into an unconstrained one, we represent each component of a particle's velocity as follows:
\begin{equation}
    \label{equ:track polynomial}
    \mathcal{P}\mathopen{}\left(t\right) \equiv v_\mathrm{p}\mathopen{}\left(t\right) = g\mathopen{}\left(t\right) + \sum_{j=1}^{n_\mathrm{c}} \eta_j \,\varphi_j\mathopen{}\left(t\right).
\end{equation}
In this expression, $g$ is an unconstrained ``free function'' that is at least once integrable and differentiable, $\eta_j$ is a projection coefficient that enforces Eq.~\eqref{equ:particle advection} for any instantiation of $g$, $\varphi_j$ is a switch function that integrates to unity within the $j$th interval and zero elsewhere, and $n_\mathrm{c} = n_k - 1$ is the number of constraints for the track, i.e., the number of intervals. All the free parameters are embedded in $g$, which we set to be a $p$th-order polynomial in time,\footnote{In practice, $p = n_k + 2$ ensures that $g$ can represent the vast majority of particle tracks to an adequate tolerance.}
\begin{equation}
    \label{equ:free function}
    g\mathopen{}\left(t\right) = \sum_{i = 0}^{p} \theta_i \,t^i.
\end{equation}
Note that the coefficients $\theta_i$ make up the trainable vector $\boldsymbol\uptheta_\mathrm{p}$. The $j$th projection coefficient is
\begin{equation}
    \label{equ:projection coefficients}
    \eta_j = \left(x_j - x_{j-1}\right) - \int_{t_{j-1}}^{t_j} g\mathopen{}\left(t\right) \mathrm{d}t,
\end{equation}
which corresponds to the integral constraint in Eq.~\eqref{equ:particle advection}, and the $j$th switch function is
\begin{equation}
    \label{equ:switch function}
    \varphi_j\mathopen{}\left(t\right) = \sum_{i = 1}^{n_\mathrm{c}} s_i\mathopen{}\left(t\right) A_{i,j},
\end{equation}
where $s_i$ is the $i$th so-called support function and $A_{i,j}$ is a weighting coefficient. Equation~\eqref{equ:switch function} ensures that the $\eta_j$ constraints are correctly applied as a function of time. This objective corresponds to the following condition,
\begin{equation}
    \int_{t_{i-1}}^{t_i} \varphi_j\mathopen{}\left(t\right) \mathrm{d}t = \left\{\begin{array}{ll}1, &\quad i = j\\ 0, &\quad i \neq j\end{array}\right.
\end{equation}
for all $i$ and $j$ in $\{1, 2, \dots, n_\mathrm{c}\}$, culminating in a linear system with three $n_\mathrm{c} \times n_\mathrm{c}$ matrices,
\begin{equation}
    \label{equ:linear constraints}
    \mathbf{SA} = \mathbf{I}.
\end{equation}
In this system, the matrix $\mathbf{S}$ has elements
\begin{equation}
    S_{i,j} = \int_{t_{i-1}}^{t_i} s_j\mathopen{}\left(t\right) \mathrm{d}t,
\end{equation}
$\mathbf{A}$ comprises the coefficients $A_{i,j}$, and $\mathbf{I}$ is the identity matrix. Support functions must be selected to ensure that $\mathbf{S}$ is non-singular, but they are otherwise arbitrary. Following the recommendation of Leake et al. \cite{Leake2022}, we employ monomial support functions: $s_j(t) = t^{j-1}$.\par

Given a set of support functions, the coefficients in $\mathbf{A}$ are computed by solving Eq.~\eqref{equ:linear constraints}. This step is independent of the form of $g$ and coefficients in $\boldsymbol\uptheta_\mathrm{p}$. Conversely, the projection coefficients, $\eta_j$, must be adjusted as a function of $\boldsymbol\uptheta_\mathrm{p}$ to preserve the integral constraints in Eq.~\eqref{equ:projection coefficients}. Using the polynomial free function in Eq.~\eqref{equ:free function}, a closed-form expression for $\eta_j$ can be obtained by substituting Eq.~\eqref{equ:free function} into Eq.~\eqref{equ:projection coefficients}. To simplify the implementation of $\mathcal{P}$, the track polynomials can be written in matrix form. First, we specify a time vector, $\boldsymbol\uptau(t) = \{t^j\}_{j=0}^{p}$; a displacement vector, $\boldsymbol\updelta = \{x_j - x_{j-1}\}_{j=1}^{n_\mathrm{c}}$; a $p + 1 \times n_\mathrm{c}$ support matrix, $\mathbf{C}$, with elements
\begin{equation}
     C_{i,j} = i^{-1}\left(t_j^i - t_{j-1}^i\right),
\end{equation}
where $t_j^i$ is the time of the $j$th measurement, $t_j$, raised to the $i$th power; and an $n_\mathrm{c} \times p + 1$ augmented weight matrix, $\hat{\mathbf{A}} = [\mathbf{A}; \mathbf{0}]^\top$. Using these elements, Eq.~\eqref{equ:track polynomial} becomes
\begin{equation}
    \label{equ:track matrix}
    v_\mathrm{p}\mathopen{}\left(t\right) = \left[\boldsymbol\uptheta_\mathrm{p}^\top \left(\mathbf{I} - \mathbf{C}\hat{\mathbf{A}}\right) + \boldsymbol\updelta^\top \hat{\mathbf{A}}\right] \boldsymbol\uptau\mathopen{}\left(t\right).
\end{equation}
The resulting function is a continuous representation of velocity that inherently satisfies Eq.~\eqref{equ:particle advection} for any $\boldsymbol\uptheta_\mathrm{p}$ and can be rapidly evaluated in a differentiable computing environment.\par

\section{Numerical demonstrations}
\label{sec:demos}
We demonstrate simultaneous flow reconstruction and particle characterization with two flows: (1)~forced isotropic turbulence seeded with bidisperse particles and (2)~supersonic flow over a cone--cylinder body. Both cases feature one-way coupled transport. For comparison, we also reconstruct the same data sets without accounting for slip, i.e., assuming the particles are ideal tracers and $\tau_\mathrm{p} = 0$~s.\par

\subsection{Forced isotropic turbulence with bidisperse particles}
\label{sec:demos:isoturb}
First, we test our method on a low-speed incompressible turbulent flow using direct numerical simulation (DNS) data from the Johns Hopkins Turbulence Database \cite{Perlman2007}. Specifically, we consider the forced homogeneous isotropic turbulence case, having a Taylor microscale Reynolds number of $Re_\uplambda = 433$. Our measurement volume is the central $128^3$-voxel region of the DNS domain, and the numerical test spans 41 frames. To mimic a real experiment, we dimensionalize the data using air for the carrier phase ($\nu_\mathrm{f} = 15$~mm$^2$/s), resulting in a domain size of $10^3$~cm$^3$, a measurement duration of 0.016~s, and a sampling rate of 2580~Hz.\par

To generate particle tracks, we simulate the advection of 50,000 spherical soda lime glass beads, whose density is fixed at 2500~kg/m$^3$. Each bead is assigned a random diameter that is drawn from one of two Gaussian distributions. The first distribution has a mean diameter of 35~$\upmu$m and standard deviation of 2~$\upmu$m; the second distribution has a mean of 70~$\upmu$m and standard deviation of 4~$\upmu$m; half the beads are drawn from the former distribution, the rest are drawn from the latter. Altogether, the particles have a volume fraction of $3\times10^{-6}$ and mass loading of $6\times10^{-3}$, placing this flow near the upper boundary of the one-way coupled transport regime \cite{Elghobashi1994, Brandt2022}. Using the fluid's Kolmogorov time scale, $\tau_\mathrm{f} = 8.2$~ms, the mean Stokes numbers of the small and large particle distributions are 1 and 5, respectively, indicating strong particle lag across the board \cite{Samimy1991}.\par

\begin{figure}[ht]
\vspace*{-0.5em}
    \centering
    \includegraphics[width=\textwidth]{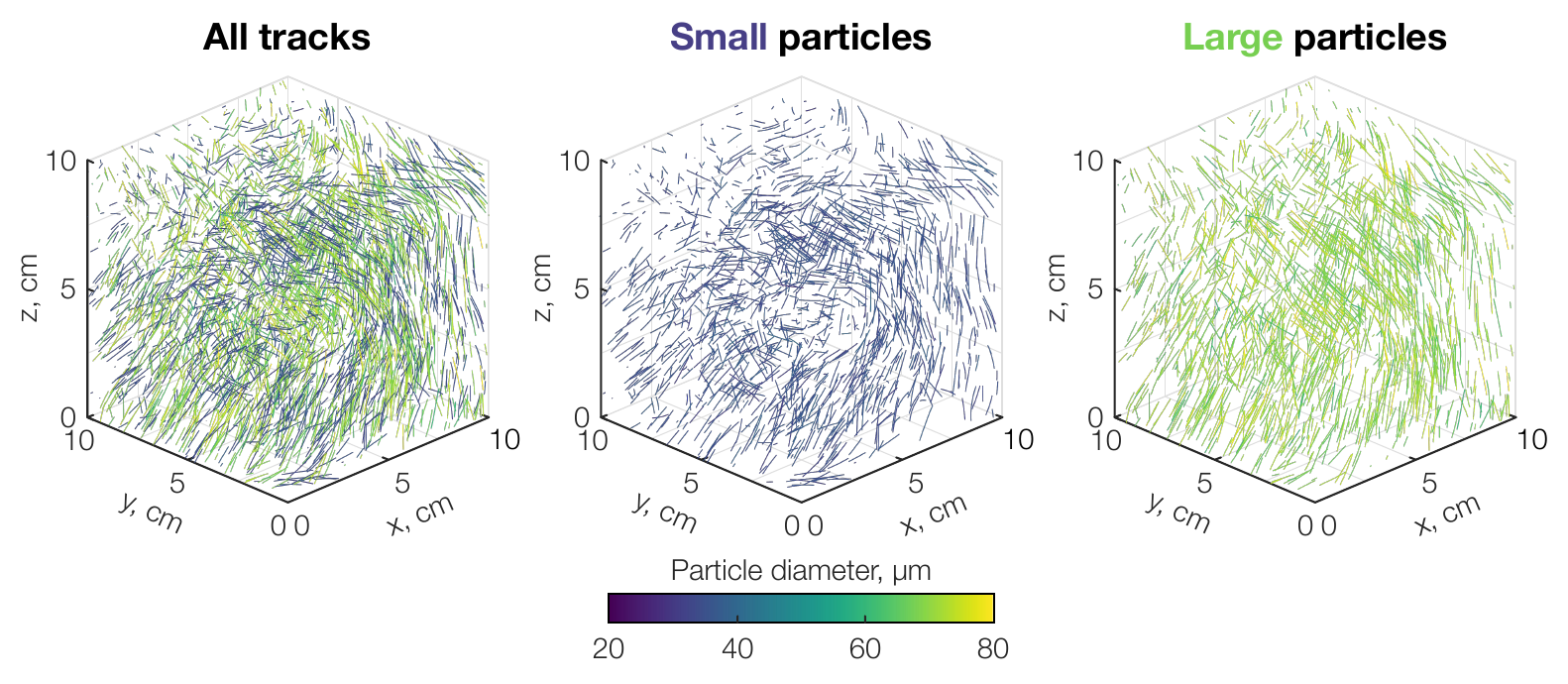}
    \caption{(left) Randomly selected tracks of bidisperse particles in isotropic turbulent flow; (middle) small particles and (right) large particles exhibit distinct behavior. Tracks are colored by the particle diameter.}
    \label{fig:isoturb tracks}
\end{figure}

The particles are randomly positioned in the cube, to start, and each particle sets off at the local flow velocity plus a settling velocity, $g \tau_\mathrm{p}$. Particle motion is modeled with the Maxey--Riley equation \cite{Maxey1997}, given in \ref{app:governing eqs:incompressible:disperse}, including steady forces like viscous drag and gravity as well as the unsteady pressure gradient and added mass terms. Following \cite{Eaton2009, Ling2013}, we omit the Basset history force, but it is readily incorporated in our framework when called for. The equation of motion is solved with a second-order Runge--Kutta scheme, wherein material derivatives are computed via seecond-order central differences, and periodic boundaries are applied to the domain walls. To ensure statistical convergence and avoid boundary effects in the tracks, we conduct our forward simulation in an enlarged domain (150 voxels) for 201 frames. Reconstructions are based on tracks contained within the central 128-voxel volume during the final 41 frames. There are about 31,000 particles in the probe volume at any given time, on average, corresponding to 0.03~ppp for a 1~MP camera. Figure~\ref{fig:isoturb tracks} depicts a random selection of 3100 tracks, which are colored by the corresponding particle diameter, $d_\mathrm{p}$. Tracks of the large and small particles are woven together in a dense cluster. The middle and right subplots of Fig.~\ref{fig:isoturb tracks} show the isolated distributions and reveal some key qualitative differences between $St \sim 1$ and $St \sim 5$ transport in this flow. While the smaller particles (purple tracks) appear to move in all directions, akin to the carrier phase, the large (chartreuse) tracks bear the mark of gravitational settling: falling in the negative $z$-direction over time. Crucially, both sets of tracks ``mask'' the underlying flow in a unique way and the particles are unlabeled, culminating in a complex reconstruction problem.\par

\begin{figure}[ht]
\vspace*{-0.5em}
    \centering
    \includegraphics[width=\textwidth]{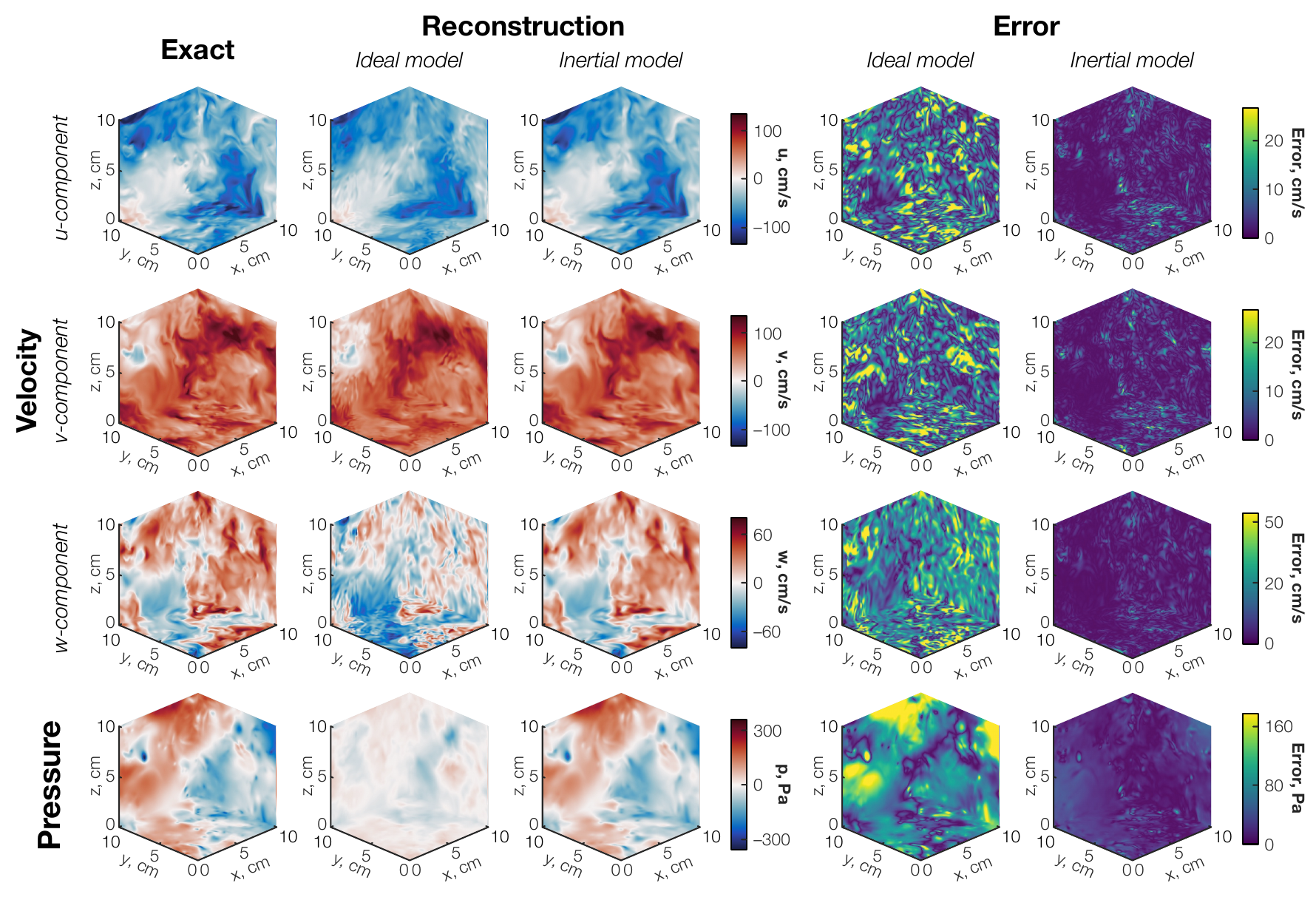}
    \caption{(left) Exact and reconstructed isotropic turbulent flow fields and (right) absolute error fields.}
    \label{fig:isoturb panel}
\end{figure}

Inertial tracks from the bidisperse particles are fed to our reconstruction algorithm, which is executed in an ``ideal tracer'' (or ``conventional'') mode and a comprehensive ``inertial'' mode. For ideal tracer reconstructions, we set $\tau_\mathrm{p}$ to zero. In our inertial reconstructions, the relaxation time of each particle is included in the trainable variable $\boldsymbol\Uppsi$, and $\tau_\mathrm{p}$ is used to determine $d_\mathrm{p}$ (see Eq.~\eqref{equ:response time:creep}). This conversion is valid because the beads have a uniform density, shape, and surface finish, so there is a unique relationship between $\tau_\mathrm{p}$ and $d_\mathrm{p}$. In the inertial case, we initialize $\tau_\mathrm{p}$ using a random diameter that is drawn from a single distribution of mean 52.5~$\upmu$m and standard deviation 4~$\upmu$m. This distribution of $d_\mathrm{p}$ has minimal overlap with either of the true size distributions.\par

Figure~\ref{fig:isoturb panel} shows cut plots of velocity and pressure from the ``ground truth'' DNS and both reconstructions. Cuts are shown at the bottom ($z = 0$~cm), rear ($y = 10$~cm), and right ($x = 10$~cm) face of the domain. (Note that we do not employ boundary conditions for these reconstructions.) While there is \textit{qualitative} agreement between the DNS velocity fields and conventional reconstructions in the $x$- and $y$-directions, significant errors are visible in the $w$-velocity and pressure fields. Acute $z$-direction errors are due to the inability of a conventional algorithm to distinguish gravitational settling and advection, and pressure cannot be recovered from inaccurate velocity data \cite{Pan2016, Faiella2021, Nie2022}. Moreover, despite the  \textit{apparent} -- i.e., visual -- agreement of the DNS and conventionally-estimated $u$- and $v$-components of $\mathbf{u}_\mathrm{f}$, the error fields on the right side of Fig.~\ref{fig:isoturb panel} indicate large numerical deviations in all the conventionally reconstructed fields. In sharp contrast, our inertial reconstructions are highly accurate, as evidenced by the dark purple (null) error maps. Normalized root-mean-square errors (NRMSEs) of the inertial reconstructions are 3.8\%, 3.3\%, and 9.7\% for the $u$-, $v$-, and $w$-components of velocity and 17.8\% for pressure. Conventional reconstructions exhibit much greater errors of 20.6\%, 20.1\%, 101.5\%, and 81.2\% for the same fields.\par

\begin{figure}[ht]
\vspace*{-0.5em}
    \centering
    \includegraphics[height=5cm]{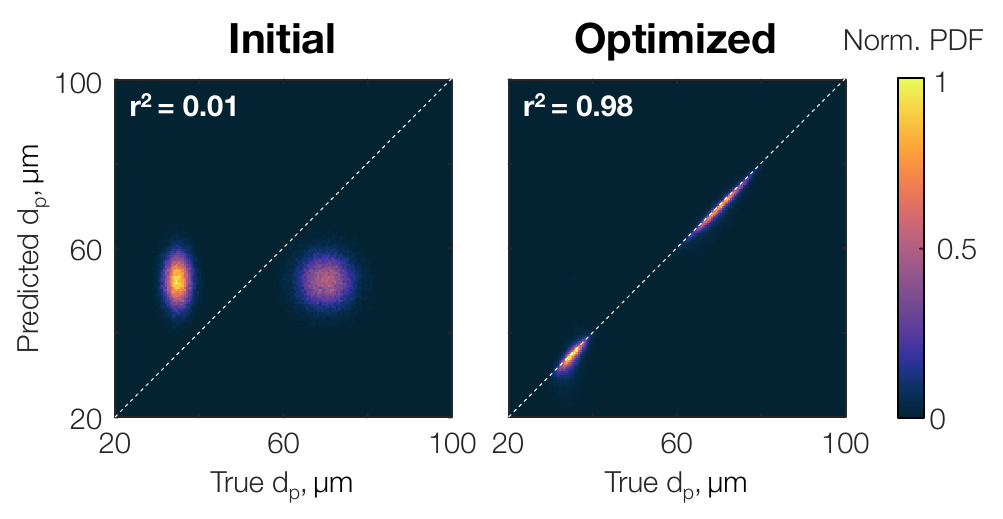}
    \caption{Implicit particle sizing results for the isotropic turbulent flow case: (left) initial and (right) optimized estimates of $d_\mathrm{p}$ compared to the true values.}
    \label{fig:isoturb PDFs}
\end{figure}

Training the inertial model yields an estimate of each particle's response time and hence diameter. Results of this inference are plotted in Fig.~\ref{fig:isoturb PDFs}, which depicts normalized joint probability density functions (PDFs) of the $d_\mathrm{p}$ estimates and exact values. There is almost perfect agreement between the true and optimized particle diameters, with a clear separation between the distinct size distributions. Consequently, the initial Pearson correlation coefficient of 0.01 rises to 0.98 after training. We emphasize that this classification does not rely on any prior knowledge of the flow states nor analysis of particle images, cf. \cite{Zhang2008, Khalitov2002}. Instead, our results are obtained from the tracks and governing physics, alone. However, in situations where additional information from another sizing technique is available, as in \cite{Huang2021, delaTorre2023}, it can be included in the optimization of $d_\mathrm{p}$ to enhance the accuracy of the flow states and particle properties, alike.\footnote{We confirmed this result through a series of tests (less challenging and not reported in this text) in which exact values of $d_\mathrm{p}$ were precisely known, resulting in better estimates of $\mathbf{u}_\mathrm{f}$ and $p_\mathrm{f}$. Statistical information about $d_\mathrm{p}$ from a calibration test is also beneficial.}\par

\subsection{High-speed flow over a cone--cylinder body}
\label{sec:demos:cone}
Second, we assess the performance of our method with a steady high-speed flow. Air flowing at Mach~2 passes over a 15$^\circ$ half-angle cone--cylinder body, resulting in an oblique shock at the nose and expansion fan past the shoulder. The inflow density and temperature are 0.55 kg/m$^3$ and 166.7 K, respectively, and the cone has a radius of 20~mm, which matches the experiment of Venkatakrishnan and Meier \cite{Venkatakrishnan2004}. We simulate the flow using the compressible axisymmetric Navier--Stokes solver in SU2~7.3.0. The computational domain has a radius of 0.15~m and length of 0.25~m. The ratio of specific heats for air is set to the standard value of 1.4. Additional details about the mesh, solver, and experimental validation are provided in \cite{Molnar2023a}.\par

Transport of 2000 spherical TiO$_{2}$ particles is simulated within the fluid domain. The particles have a fixed density of 3500~kg/m$^3$ and a random diameter that we draw from a Gaussian distribution of mean 1.5~$\upmu$m and standard deviation $0.2$~$\upmu$m.\footnote{This is a realistic range for TiO$_{2}$ seed used in high-speed PIV \cite{Melling1997}. Real tracers are also subject to density variations caused by agglomeration and humidity \cite{Williams2015}, but we do not consider these effects at this stage.} The average $\tau_\mathrm{p}$ of these particles is about 38~$\upmu$s, corresponding to strong particle inertia, and the volume fraction and mass loading are $10^{-9}$ and $10^{-6}$, respectively, resulting in one-way coupled dynamics. Particles are injected into the domain at the inlet, starting their journey at the free-stream velocity, and advected downstream via Eq.~\eqref{equ:particle motion}. This equation is solved using the numerical scheme described in Sec.~\ref{sec:demos:isoturb}. Particle positions are saved at a rate of 0.5~MHz, which is within range of several state-of-the-art high-speed PIV and LPT setups \cite{Beresh2020, Beresh2021, Manovski2021}. Similar to the incompressible case, we advect the particles for 201 timesteps and retain the final eight frames to reconstruct the flow. We opt for eight frames to match the limitations of cutting-edge illumination and imaging hardware \cite{Bordoloi2018, Lin2023}. Admittedly, four-pulse LPT is more technically mature and has been recently demonstrated on a subsonic jet \cite{Manovski2021} and turbulent boundary layer \cite{Novara2016}. But shorter tracks convey less information about the flow and particle dynamics, leading to less accurate acceleration data and a more challenging reconstruction problem \cite{Schroder2023}. The relationship between reconstruction accuracy and track length merits further investigation. It should also be noted that MHz-rate PIV and LPT experiments usually have a measurement domain length scale of order 1~cm, owing to the reduced sensor resolution of most high-speed cameras when operated at their maximum acquisition rate \cite{Beresh2020}. The present demonstration features a length scale of 10~cm to highlight a range of flow features, but our method is equally applicable to smaller domains.\par

\begin{figure}[ht]
    \vspace*{-0.5em}
    \centering
    \subcaptionbox{\label{fig:cone tracks:tracks}}{\includegraphics[height=5cm]{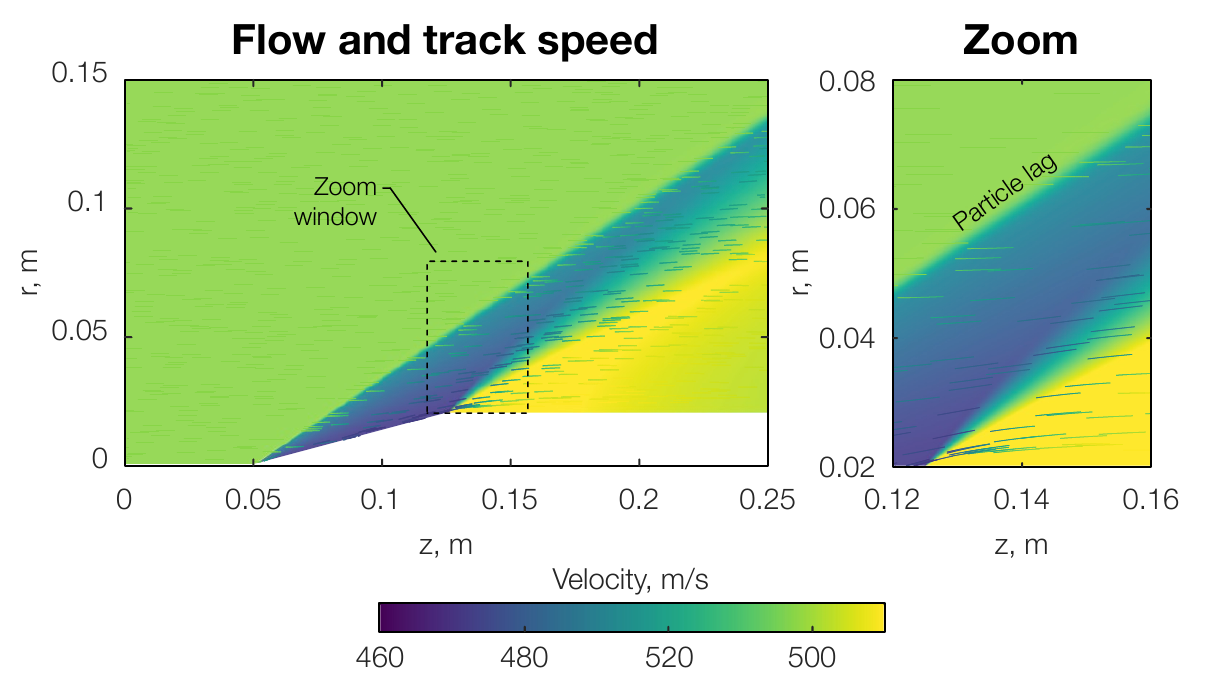}}\quad
    \subcaptionbox{\label{fig:cone tracks:response time}}{\includegraphics[height=5cm]{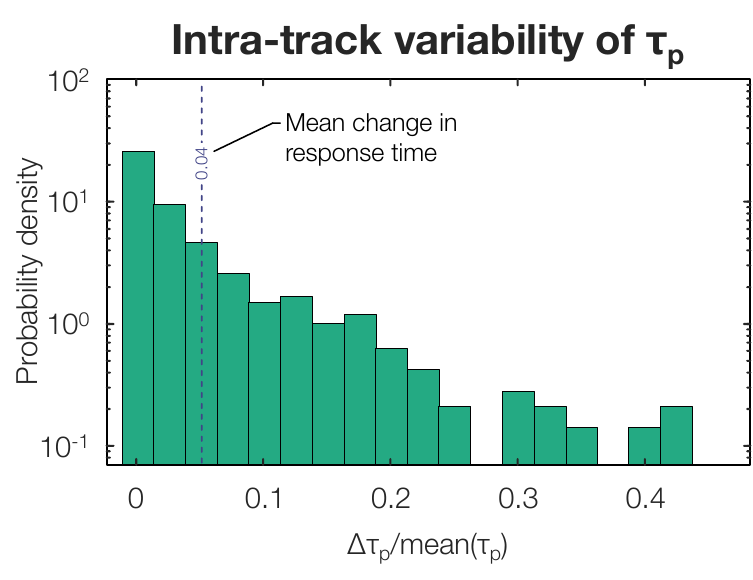}}
    \caption{(a) Tracks of TiO$_2$ particles in supersonic flow and (b) PDF of the normalized intra-track range of $\tau_\mathrm{p}$. Tracks are colored by the local particle speed and the background is colored by the flow speed; hence, tracks are only visible when there is slip.}
    \label{fig:cone tracks}
\end{figure}

Figure~\ref{fig:cone tracks:tracks} shows simulated tracks from the cone--cylinder flow, colored by the local particle speed. The background color indicates the flow speed, so the particle tracks are only visible at locations with a slip velocity between the phases. Slip can be observed in the aft-shock region and through the expansion fan, wherein particles lag the air due to strong deceleration and acceleration, respectively. Notably, large gradients in the viscosity, density, and speed of sound of a shocked carrier phase have a leading-order effect on the particle dynamics \cite{Williams2014, Williams2015}. As a result, the response time of a particle in supersonic flow is an unsteady quantity that may change throughout the measurement domain. Figure~\ref{fig:cone tracks:response time} shows PDFs of the normalized dynamic range of $\tau_\mathrm{p}$ for particles that travel through the shock or expansion fan. Note that this figure shows the \textit{intra}-track variablility of $\tau_\mathrm{p}$, not the intertrack variability. Particles included in Fig.~\ref{fig:cone tracks:response time} are identified via a particle Mach number, $Ma_\mathrm{p}$, greater than 0.01. We exclude particles upstream of the shock due to their zero-slip initialization and constant $\tau_\mathrm{p}$. Per Fig.~\ref{fig:cone tracks:response time}, $\tau_\mathrm{p}$ changes by an average of 4\% along a single 8~mm track, with changes up to 40\% experienced by particles that cross the shock wave. The transient nature of $\tau_\mathrm{p}$ is central to particle transport in high-speed flow and has been neglected in the correction methods developed for high-speed PIV, as reviewed in \ref{app:inertial particle velocimetry:high-speed}.\par

\begin{figure}[t]
\vspace*{-0.5em}
    \centering
    \includegraphics[width=\textwidth]{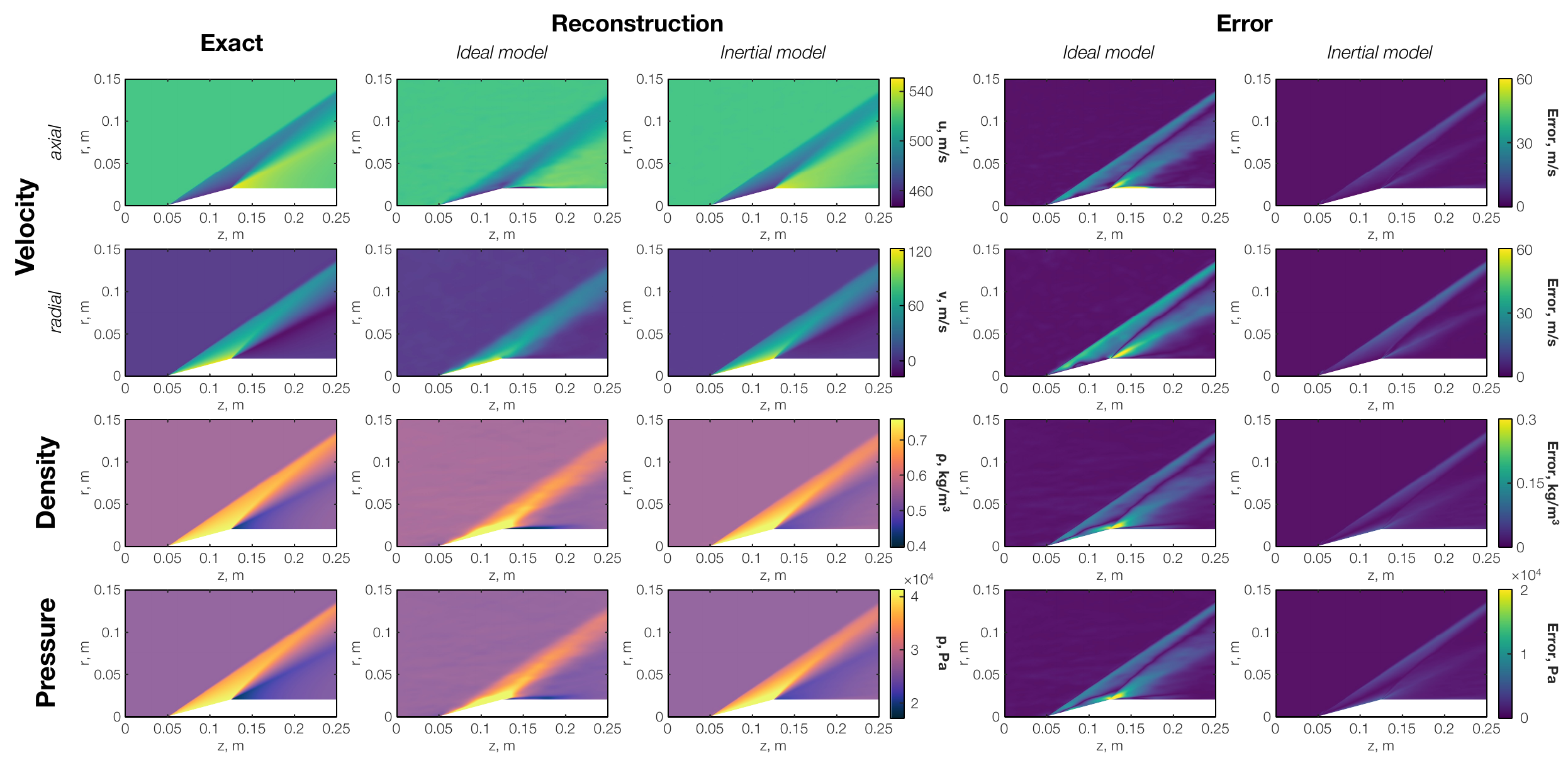}
    \caption{(left) Exact and reconstructed cone--cylinder flow fields and (right) absolute error fields.}
    \label{fig:cone panel}
\end{figure}

\begin{figure}[ht]
\vspace*{-0.5em}
    \centering
    \includegraphics[height=5cm]{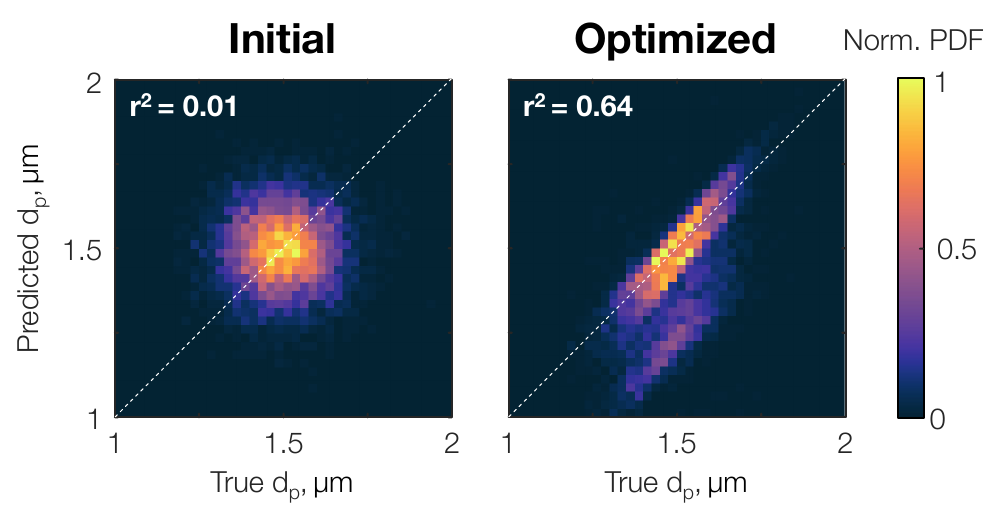}
    \caption{Implicit particle sizing results for the cone--cylinder flow case: (left) initial and (right) optimized estimates of $d_\mathrm{p}$ compared to the true values.}
    \label{fig:cone PDFs}
\end{figure}

Reconstructed velocity, density, and pressure fields of the cone--cylinder flow are plotted in Fig.~\ref{fig:cone panel}. Once again, results are presented for ``ideal tracer'' and ``inertial'' reconstructions. We observe that the conventional reconstruction blurs the shock interface, which is consistent with the literature \cite{Samimy1991, Ragni2011}. At the same time, the reconstruction contains high-frequency artifacts associated with changes in $\tau_\mathrm{p}$, which varies with $d_\mathrm{p}$ and the local flow state. Response time variations manifest as spurious velocity gradients when $\tau_\mathrm{p}$ is assumed to be zero. Previous PIV measurements did not exhibit these artifacts due to the smoothing effect of cross-correlation algorithms \cite{Ragni2011} (which also suppress bona fide turbulent fluctuations \cite{Williams2014}). By contrast, our method accurately resolves the shock structure and latent fields. On the right side of Fig.~\ref{fig:cone panel}, we show absolute errors of the reconstructions, confirming the superiority of our algorithm. NRMSEs of our inertial reconstructions are 0.5\%, 5.5\%, 5.2\%, and 26\% for the axial velocity, radial velocity, density, and pressure fields, respectively, compared to 7.4\%, 27.2\%, 15.5\%, and 75.3\% for the conventional method. Lastly, Fig.~\ref{fig:cone PDFs} plots the results of implicit particle sizing in supersonic flow. Optimized values of $d_\mathrm{p}$ roughly predict the true diameters, with some slight deviations. Since the reconstructed flow fields are highly accurate, errors in $d_\mathrm{p}$ may be explained by a multiplicity of weak solutions to Eqs.~\eqref{equ:Maxey-Riley} and \eqref{equ:Loth} when $d_\mathrm{p}$ is unknown.\par

\section{Conclusions}
\label{sec:conclusions}
We present a novel method for reconstructing flow states and estimating the physical properties of tracer particles from LPT data. Inertial particle transport is an important source of error in many PIV and LPT experiments, and we account for this effect via the equations that govern disperse multiphase flow. We solve these equations in a weak sense by optimizing neural flow states and polynomial particle tracks that embed the LPT data as a hard constraint. Our algorithm is tested numerically on incompressible isotropic turbulence and steady supersonic flow over a cone--cylinder test article. Both data sets feature inertial particles whose size is estimated during the reconstruction procedure. Our method significantly improves the accuracy of flow states determined from inertial tracks, as compared to the conventional approach in which the particles are assumed to faithfully follow the flow. Moreover, we predict the size of each particle with acceptable accuracy using the track data and governing physics, alone (as opposed to optical sizing). In future work, we will (1)~utilize developments from \cite{Zhou2023} to account for localization and tracking errors in our reconstructions; (2)~characterize the accuracy and resolution limits of our method as a function of the seeding density, measurement and flow scales, and particle inertia; and (3)~apply the technique to experimental data sets.\par

\appendix
\renewcommand{\thesection}{Appendix \Alph{section}}

\section{Inertial transport in particle-based velocimetry}
\label{app:inertial particle velocimetry}

\subsection{High-speed flows}
\label{app:inertial particle velocimetry:high-speed}
Interpreting high-speed LPT and PIV data has long been stymied by inertial transport \cite{Ross1994}. Additionally, there is evidence of finite-$St$ effects in relatively low-speed experiments. For example, Wolf et al. \cite{Wolf2019} performed LPT measurements of a helicopter wake flow with a blade-tip Mach number of 0.3. Despite using 350~$\upmu$m helium-filled soap bubble tracers -- a common choice for aerodynamic flows, having a \textit{nominal} Stokes number around 0.1 in Wolf's test -- the blade-tip vortices in \cite{Wolf2019} were devoid of bubbles, which is indicative of inertial clustering. This result suggests that the local flow time scale, length scale, or both fell within range of the corresponding particle scale(s), leading to slip or even geometric effects. These phenomena could compromise reconstructions, and empty vortex cores like those in \cite{Wolf2019, Nair2023} raise concerns about the reliability of tracers in LPT measurements of turbulent aerodynamic flows. In faster flows, such as compressible wakes and supersonic turbulent boundary layers, inertial transport is much more pronounced. Urban and Mungal \cite{Urban2001} applied PIV measurement to a compressible mixing layer at a convective  Mach number of 0.63. They observed preferential sampling of the Al$_2$O$_3$ tracer particles ($St \approx 0.5$) in large-scale mixing vortices. A numerical study of a similar flow, conducted by Samimy and Lele \cite{Samimy1991}, showed a linear increase in slip velocity (up to 10\%) with $St$ (up to 1). Brooks et al. \cite{Brooks2018} performed PIV measurements of a Mach~3 turbulent boundary layer and found that particles with $St > 1$ effectively filter the observed velocity fluctuations near the wall, which can produce Reynolds stress errors up to 30\%. This damping effect is especially pronounced in the wall-normal direction, which is a challenge for model verification \cite{Bernardini2011, Bross2021}. Further, there is a wealth of experimental evidence that inertial ``filtering'' flattens out the turbulent kinetic energy spectrum \cite{Brooks2018, Lowe2014, Williams2014, Williams2018, Aultman2022}. Recent simulations have corroborated these measurements, e.g., Williams et al. \cite{Williams2015} predicted Stokes numbers around two for near-wall particles in a Mach~7 turbulent boundary layer, suggesting strong lag. These effects make it difficult to use PIV and LPT for experimental investigation of high-speed flows and model validation of the same.\par

Shock waves play an important role in the dynamics of supersonic wall-bounded flows and produce severe particle lag. Particles traversing a shock slow down exponentially to the post-shock flow speed \cite{Schrijer2007, Ragni2011}. This decay may be characterized by $\tau_\mathrm{p}$ and the particle's relaxation distance, $l_\mathrm{p}$, defined as the distance traveled by a particle during $\tau_\mathrm{p}$ \cite{Ragni2011}. Well-controlled LPT and PIV measurements of TiO$_2$ and Al$_2$O$_3$ particles passing through oblique shocks yielded estimates of $\tau_\mathrm{p}$ between 2 and 20~$\upmu$s and $l_\mathrm{p}$ close to 1~mm \cite{Ragni2011, Scarano2003, Urban2001}. This distance considerably exceeds the characteristic width of a typical shock, which is on the order of microns. As a result, particle tracks shift the apparent shock location and blur the velocity jump, causing biased errors in the range of 20--35\% \cite{Glazyrin2015, Koroteeva2018}. Despite these lag-induced limitations, the step change of flow conditions across a shock presents an opportunity to benchmark unknown particle properties like size and density. The particle speed decay can be modeled by an equation of motion that features a quasi-steady drag force \cite{Tedeschi1999, Williams2015}, as discussed in \ref{app:governing eqs:compressible:disperse}. This force depends not only on the particle properties, but also on local values of the carrier phase velocity, density, and temperature, which vary throughout the post-shock region. Given precise knowledge of the upstream and downstream conditions, one can calibrate $\tau_\mathrm{p}$ for each particle by fitting the transport model to the tracks \cite{Melling1997, Scarano2003}; one can subsequently infer particle size and density distributions, as done by Williams et al. \cite{Williams2015}. We note that the pertinent drag models are highly nonlinear in these particle and fluid properties and the response time is an unsteady quantity. In spite of this, $\tau_\mathrm{p}$ is generally treated as a constant along a track \cite{Williams2015}. Uncertainty about flow conditions and the use of a fixed value of $\tau_\mathrm{p}$ for each particle thus lead to large errors in estimates of $d_\mathrm{p}$ and the particle density, $\rho_\mathrm{p}$.\par

One way to ameliorate inertial effects is to correct the \textit{apparent} velocity field, obtained from the particle images via cross-correlation and averaging, in a post-processing step \cite{Koike2007}. Existing corrections have been developed for PIV measurements of planar shocks \cite{Schrijer2007, Boiko2013} and supersonic jets \cite{Boiko2015}. These algorithms rely upon the transport model and (potentially inaccurate) particle properties estimated from a controlled experiment, as described above, which complicates the experimental workflow. Moreover, the corrections are only valid for steady or time-averaged data, and they feature a smooth ``fictitious'' particle motion field that is extracted from the images with a cross-correlation algorithm. Extending such methods to the unsteady turbulent case is challenging because inertial particles in supersonic flow exhibit chaotic motion with a large dynamic range, and they cut across one another's path, resulting in cross-correlation errors and inaccurate kinematics. Our method addresses this issue by simultaneously estimating the particle properties and reconstructing the flow states via comprehensive disperse and carrier phase transport models.\par

\subsection{Inherently multiphase flows}
\label{app:inertial particle velocimetry:natural}
High-$St$ particle transport is ubiquitous in natural and engineering flows, including sea-bed mixing, dust storms, snowfall, sprays, and bubbly flows. These and other such multiphase scenarios present an opportunity for passive LPT, provided that inertial effects are included in the reconstruction procedure. Examples include sand-based LPT tests of sediment mixing, which reveal long, persistent sediment streaks at the bottom of a water channel \cite{Righetti2004, Muste2005}. Such streaks could be a manifestation of inertial clustering caused by wall coherent structures. Although weak streaks of light (passive) particles have been seen in some conditions \cite{Rashidi1990}, heavy inertial particles tend to form denser and more prominent streaks \cite{Muste2005}. Simultaneous imaging of neutrally-buoyant tracers may be performed to resolve the water velocity field. However, mixed tracer--inertial particle tests currently require image-based particle classification, which is subject to large uncertainties \cite{Zhang2008, Khalitov2002}. Another natural experiment involves field-scale LPT and PIV measurements of atmospheric turbulence using images of snow, wherein strong gravitational settling and preferential sampling are observed \cite{Bristow2023, Li2021}. Stokes numbers in these tests were estimated to range from 0.1 to 1, meaning the snowflake trajectories did not directly indicate the local carrier phase velocity. Turbulent bubbly flow is another case where natural particles can be leveraged for LPT, although bubbles may fail as tracers due to non-ideal effects like buoyancy \cite{Tan2020}. Furthermore, given a large volume fraction of bubbles, the flow may sustain two-way coupling of the disperse and carrier phases \cite{Brandt2022}, rendering flow reconstruction based on bubble tracking a formidable task. Our reconstruction technique has the potential to address these challenges and facilitate the use of natural particles for accurate flow state estimation.\par

\section{Governing equations}
\label{app:governing eqs}
This appendix contains governing equations for both of the particle-laden flows considered in this work. Since the particle volume fraction and mass loading are low in PIV and LPT, particle--fluid interactions may be modeled via one-way coupling. In other words, the disperse phase is passive. Particles are treated as rigid spheres, for simplicity, and they are characterized by their density and hydrodynamic diameter. The particle dynamics are governed by an equation of motion that contains the leading hydrodynamic and body forces.\par

\subsection{Unsteady 3D incompressible flow}
\label{app:governing eqs:incompressible}

\subsubsection{Carrier phase}
\label{app:governing eqs:incompressible:carrier}
Forced isotropic turbulence in Sec.~\ref{sec:demos:isoturb} is governed by the incompressible 3D continuity and momentum equations,
\begin{subequations}
    \begin{align}
        \nabla \cdot \mathbf{u}_\mathrm{f} &= 0 \quad\text{and} \label{equ:isoturb equations:continuity}\\
        \frac{\partial \mathbf{u}_\mathrm{f}}{\partial t} +  \mathbf{u}_\mathrm{f} \cdot \nabla \mathbf{u}_\mathrm{f} &= -\frac{1}{\rho_\mathrm{f}}\nabla p_\mathrm{f} + \nu_\mathrm{f}\nabla^2\mathbf{u}_\mathrm{f} + \frac{1}{\rho_\mathrm{f}}\mathbf{f}, \label{equ:isoturb equations:momentum}
    \end{align}
    \label{equ:isoturb equations}%
\end{subequations}
where $\mathbf{f}$ is a forcing term that generates stationary turbulence \cite{Rosales2005},
\begin{equation}
    \mathbf{f} = \frac{\varepsilon}{3 \,u^{2}_\mathrm{rms}} \mathbf{u}_{\mathrm{f}},
    \label{equ:isoturb equations:forcing}
\end{equation}
$\varepsilon$ is the mean energy dissipation rate, and $u_\mathrm{rms}$ is a root-mean-square velocity. Note that $\mathbf{u}_\mathrm{f}$ is a 3D vector in Eq.~\eqref{equ:isoturb equations} and the del operator, $\nabla$, is defined in Cartesian coordinates. For the isotropic turbulent flow case, the vector $\mathbf{e}_\mathrm{f}$ in Eq.~\eqref{equ:objective loss:flow} contains the residual from each component of Eq.~\eqref{equ:isoturb equations}.\par

\subsubsection{Disperse phase}
\label{app:governing eqs:incompressible:disperse}
Small spherical particles moving in (locally) uniform flow are subject to inertial and viscous effects. This is an apt description of the soda lime glass beads in Sec.~\ref{sec:demos:isoturb}. Inertial transport regimes are often delineated in terms of the particle Reynolds number, defined with respect to a particle length scale, slip velocity, and fluid viscosity,
\begin{equation}
    Re_\mathrm{p} = \frac{d_\mathrm{p} \overbrace{\left|\mathbf{u}_\mathrm{f} - \mathbf{v}_\mathrm{p}\right|}^\text{slip}}{\nu_\mathrm{f}}.
    \label{equ:Reynolds}
\end{equation}
Slip, i.e., lagging or ballistic particle motion, is characterized in terms of a response time,
\begin{equation}
    \tau_\mathrm{p} = \frac{4}{3 \,C_\mathrm{D} \,Re_\mathrm{p}} \frac{\rho_\mathrm{p} \,d_\mathrm{p}^2}{\rho_\mathrm{f} \,\nu_\mathrm{f}},
    \label{equ:response time:general}
\end{equation}
where $C_\mathrm{D}$ is the drag coefficient. Creeping flow occurs when $Re_\mathrm{p} \ll 1$, which yields a particle Reynolds number of $24/C_\mathrm{D}$ such that
\begin{equation}
    \tau_\mathrm{p} = \frac{1}{18} \frac{\rho_\mathrm{p} \,d_\mathrm{p}^2}{\rho_\mathrm{f} \,\nu_\mathrm{f}}.
    \label{equ:response time:creep}
\end{equation}
The small and large particle distributions in Sec.~\ref{sec:demos:isoturb} have a mean $Re_\mathrm{p}$ of 0.09 and 0.2, meaning that Eq.~\eqref{equ:response time:creep} is valid for that flow case.\par

When particles are smaller than the relevant hydrodynamic length scale, their motion is governed by the Maxey--Riley equation \cite{Maxey1997},
\begin{align}
    \frac{\mathrm{d} \mathbf{v}_\mathrm{p}}{\mathrm{d} t} =&
    \overbrace{\frac{\mathbf{u}_\mathrm{f} - \mathbf{v}_\mathrm{p}}{\tau_\mathrm{p}}}^\text{I} +
    \overbrace{\frac{\rho_\mathrm{f}}{\rho_\mathrm{p}} \frac{ \mathrm{D}\mathbf{u}_\mathrm{f}}{\mathrm{D}t}}^\text{II} +
    \overbrace{\frac{1}{2} \frac{\rho_\mathrm{f}}{\rho_\mathrm{p}} \left(\frac{ \mathrm{D}\mathbf{u}_\mathrm{f}}{\mathrm{D}t} - \frac{\mathrm{d}\mathbf{v}_\mathrm{p}}{\mathrm{d}t} \right)}^\text{III} +\\
    &\quad \underbrace{\sqrt{\frac{9}{2\pi} \frac{\rho_\mathrm{f}}{\rho_\mathrm{p} \tau_\mathrm{p}}} \int_{-\infty}^t \frac{1}{\sqrt{t-\tau}} \frac{\mathrm{d} (\mathbf{u}_\mathrm{f} - \mathbf{v}_\mathrm{p})}{\mathrm{d} \tau} \mathrm{d}\tau}_\text{IV} +
    \underbrace{\left(1 - \frac{\rho_\mathrm{f}}{\rho_\mathrm{p}}\right)\mathbf{g}}_\text{V},
    \label{equ:Maxey-Riley}
\end{align}
where $\mathrm{D}/\mathrm{D}t$ and $\mathrm{d}/\mathrm{d}t$ are the total derivatives defined with respect to a fluid parcel and particle, respectively. Terms on the right side of this expression represent the (I)~Stokes (a.k.a. quasi-steady viscous) drag, (II)~pressure gradient, (III)~added mass, (IV)~Basset history, and (V)~gravitational/buoyancy forces. Two simplifications are made based on the small and heavy particles used in our test. First, given the large particle-to-fluid density ratio, $\rho_\mathrm{p}/\rho_\mathrm{f} \sim \mathcal{O}(10^3)$, we neglect the Basset history force \cite{Eaton2009, Ling2013} in our forward simulation. In our reconstructions, we further neglect the pressure gradient and added mass forces, because they are three orders of magnitude smaller than Stokes drag, and we omit the contribution of buoyancy to (V). Second, due to the minute size of our particles, we neglect finite-size effects such as the Fax{\'e}n correction and Saffman lift \cite{Maxey1997}. The Kolmogorov length scale in Sec.~\ref{sec:demos:isoturb} is 350~$\upmu$m, which considerably exceeds the maximum particle size and supports our use of the Maxey--Riley equation. Hence, for the isotropic turbulent flow case, the vector $\mathbf{e}_\mathrm{p}^k$ in Eq.~\eqref{equ:objective loss:particle} contains the residual from each component of Eq.~\eqref{equ:Maxey-Riley}, excluding terms (II)--(IV), for the $k$th particle.\par

\subsection{Steady axisymmetric compressible flow}
\label{app:governing eqs:compressible}

\subsubsection{Carrier phase}
\label{app:governing eqs:compressible:carrier}
Cone--cylinder flow in Sec.~\ref{sec:demos:cone} is governed by the steady axisymmetric compressible continuity, momentum, and energy equations,
\begin{subequations}
    \begin{align}
        \nabla \cdot \left(\rho_\mathrm{f} \,\mathbf{u}_\mathrm{f}\right) &= 0, \label{equ:cone equations:continuity}\\
        \nabla \cdot \left(\rho_\mathrm{f} \,\mathbf{u}_\mathrm{f} \,\mathbf{u}_\mathrm{f}^\top\right) &= -\nabla p_\mathrm{f} +
        \nabla \cdot \left[\mu_\mathrm{f} \left(\nabla\mathbf{u}_\mathrm{f} + \nabla\mathbf{u}_\mathrm{f}^\top \right) -
        \frac{2}{3} \mu_\mathrm{f} \left(\nabla \cdot \mathbf{u}_\mathrm{f}\right) \mathbf{I}\right], \quad\text{and} \label{equ:cone equations:momentum}\\
        \nabla \cdot \left[\left(\rho_\mathrm{f}E_\mathrm{f} + p\right)\mathbf{u}_\mathrm{f}\right] &= \nabla \cdot \left(\kappa_\mathrm{f} \nabla T_\mathrm{f}\right) +
        \nabla \cdot \left\{\left[\mu_\mathrm{f} \left(\nabla\mathbf{u}_\mathrm{f} + \nabla\mathbf{u}_\mathrm{f}^\top \right) -
        \frac{2}{3} \mu_\mathrm{f} \left(\nabla \cdot \mathbf{u}_\mathrm{f}\right) \mathbf{I}\right] \cdot \mathbf{u}_\mathrm{f}\right\}. \label{equ:cone equations:energy}
    \end{align}
    \label{equ:cone equations}%
\end{subequations}
In these expressions, $\mathbf{u}_\mathrm{f}$ contains the axial and radial components of velocity; the 2D del operator is defined for a system of axial and radial coordinates, accordingly; and $E_\mathrm{f}$, $T_\mathrm{f}$, and $\kappa_\mathrm{f}$ are the total energy, temperature, and thermal conductivity of the carrier phase. Note that $T_\mathrm{f}$ is determined by the fluid's local total energy and velocity magnitude, and we calculate the dynamic viscosity and thermal conductivity via Sutherland's Law \cite{Anderson1990}. Equation~\eqref{equ:cone equations} contains four equations and five unknowns and must be closed with an equation of state,
\begin{equation}
    p_\mathrm{f} = \left(\gamma_\mathrm{f} - 1\right) \rho_\mathrm{f} \underbrace{\left(E_\mathrm{f} - \frac{1}{2} \mathbf{u}_\mathrm{f} \cdot \mathbf{u}_\mathrm{f}\right)}_{C_\mathrm{v} \,T_\mathrm{f}},
\end{equation}
where $\gamma_\mathrm{f}$ is the ratio of specific heats in the carrier phase and $C_\mathrm{v}$ is the specific heat by constant volume. For the cone--cylinder flow case, the vector $\mathbf{e}_\mathrm{f}$ in Eq.~\eqref{equ:objective loss:flow} contains the residual from each component of Eq.~\eqref{equ:cone equations}.\par

\subsubsection{Disperse phase}
\label{app:governing eqs:compressible:disperse}
Particle drag in high-speed flow is sensitive to compressibility and rarefaction effects in addition to viscous action by the carrier phase. Compressibility effects scale nonlinearly with the particle Mach number,
\begin{equation}
    Ma_\mathrm{p} = \frac{|\mathbf{u}_\mathrm{f} - \mathbf{v}_\mathrm{p}|}{\sqrt{\gamma_\mathrm{f} R_\mathrm{f} T_\mathrm{f}}}, 
    \label{equ:Mach}
\end{equation}
where $R_\mathrm{f}$ is the carrier phase gas constant. Rarefaction is a function of the mean free path in the carrier phase, $\lambda_\mathrm{f}$, and characteristic flow length scale, which is taken to be the particle diameter when $Re_\mathrm{p}$ is low. The ratio of $\lambda$ to $d_\mathrm{p}$ is the particle Knudsen number, $Kn_\mathrm{p}$, which may be expressed in terms of $Re_\mathrm{p}$ and $Ma_\mathrm{p}$,
\begin{equation}
    Kn_\mathrm{p} = \frac{\lambda_\mathrm{f}}{d_\mathrm{p}} = \frac{Ma_\mathrm{p}}{Re_\mathrm{p}} \sqrt{\frac{\pi \gamma_\mathrm{f}}{2}},
    \label{equ:Knudsen}
\end{equation}
wherein the right-most side makes use of the ideal gas law. Drag laws for compressible particle-laden flow can thus be specified using any two of $Re_\mathrm{p}$, $Ma_\mathrm{p}$, and $Kn_\mathrm{p}$.\par

Tracer particles in supersonic flow are well modeled as solid spheres moving in a fluid of infinite extent and subject to quasi-steady drag \cite{Williams2014}. The full Maxey--Riley equation can be rewritten to account for compressibility and rarefaction effects, per Capecelatro and Wagner \cite{Capecelatro2023}, but the pressure gradient, added mass, Basset history, and body forces have a negligible effect on the trajectory of PIV/LPT seed particles in high-speed flow \cite{Melling1997, Ragni2011}. Hence, Eq.~\eqref{equ:Maxey-Riley} reduces to
\begin{equation}
    \frac{\mathrm{d} \mathbf{v}_\mathrm{p}}{\mathrm{d} t} = \frac{\mathbf{u}_\mathrm{f} - \mathbf{v}_\mathrm{p}}{\tau_\mathrm{p}},
    \label{equ:particle motion}
\end{equation}
where $\tau_\mathrm{p}$ is given by Eq.~\eqref{equ:response time:general}. For the cone--cylinder flow case, $\mathbf{u}_\mathrm{f}$ and $\mathbf{v}_\mathrm{p}$ are 2D vectors with a radial and axial component, and the vector $\mathbf{e}_\mathrm{p}^k$ in Eq.~\eqref{equ:objective loss:particle} contains the residual from both components of Eq.~\eqref{equ:particle motion} for the $k$th particle. This expression is valid for small particles ($d_\mathrm{p} \sim 1$~$\upmu$m) in the high-density-ratio limit ($\rho_\mathrm{p}/\rho_\mathrm{f} \gg 1$). We note that the pressure gradient, added mass, and Basset history forces can eclipse Stokes drag while a particle transects a shock. However, the integrated contribution of the former three forces to the particle's trajectory is negligible in the high-density-ratio limit \cite{Thomas1992, Parmar2009, Capecelatro2023}.\par

\begin{figure}[b]
\vspace*{-0.5em}
    \centering
    \includegraphics[height=5cm]{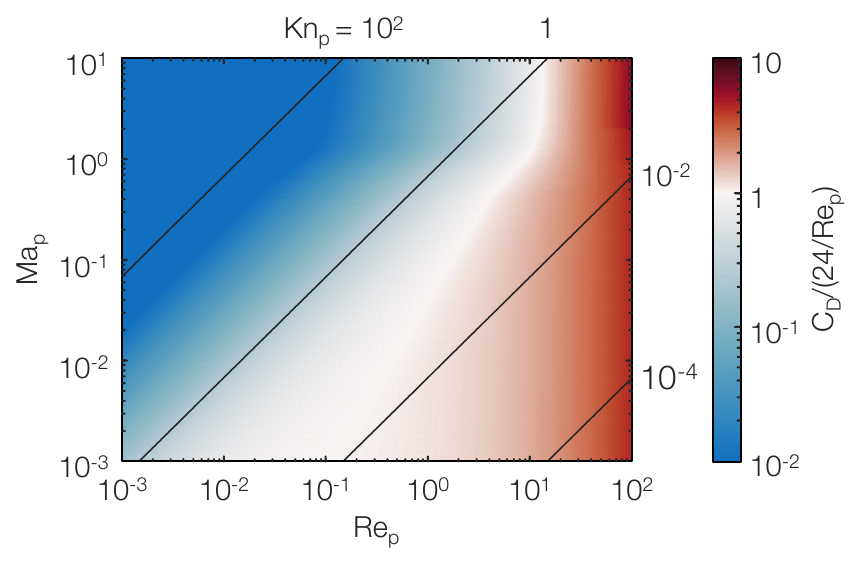}
    \caption{Loth drag model normalized by Stokes drag; black lines are isocontours of $Kn_\mathrm{p}$.}
    \label{fig:Loth}
\end{figure}

Loth \cite{Loth2008} put forth a comprehensive Stokes drag model for compressible particle-laden flows. His model is used to calculate $C_\mathrm{D}$ in terms of a particle's Reynolds and Mach numbers and hence determine $\tau_\mathrm{p}$ via Eq.~\eqref{equ:response time:general}. The overall drag model is given by
\begin{equation}
    C_\mathrm{D} = \left\{\begin{array}{ll}\dfrac{C_\mathrm{D, Kn, Re}}{1 + Ma_\mathrm{p}^4} + \dfrac{Ma_\mathrm{p}^4 \,C_\mathrm{D, f_M, Re}}{1+Ma_\mathrm{p}^4}, &\quad Re_\mathrm{p} < 45 \\
    \frac{24}{Re_\mathrm{p}} \left[1 + 0.15 Re_\mathrm{p}^{0.687}\right] H_\mathrm{M} + \dfrac{0.42 \,C_\mathrm{M}}{1 + \frac{42,500 \,G_\mathrm{M}}{Re_\mathrm{p}^{1.16}}}, &\quad Re_\mathrm{p} > 45\end{array}\right..
    \label{equ:Loth}
\end{equation}
While we formally include all of Eq.~\eqref{equ:Loth} in our forward simulation, only the rarefaction-dominated terms ($Re_\mathrm{p} < 45$) are needed for our reconstructions as the maximum $Re_\mathrm{p}$ in the cone--cylinder flow is about 16. Figure~\ref{fig:Loth} presents a map of $C_\mathrm{D}$ versus $Re_\mathrm{p}$ and $Ma_\mathrm{p}$, where $C_\mathrm{D}$ has been normalized by the Stokes drag coefficient. Loth's model spans two regimes that break across $Re_\mathrm{p} = 45$. Below this threshold is a rarefaction-dominated regime, in which $C_\mathrm{D}$ varies primarily with $Kn_\mathrm{p}$, above it is a compression-dominated regime, wherein $Ma_\mathrm{p}$ has a controlling effect. Both limiting behaviors are evident in Fig.~\ref{fig:Loth}: at low Reynolds numbers, isocontours of $C_\mathrm{D}$ are aligned with those of $Kn_\mathrm{d}$; towards higher values of $Re_\mathrm{p}$, the gradient of $C_\mathrm{D} \,Re_\mathrm{p}/24$ bends towards the Mach axis.\par

Returning to Eq.~\eqref{equ:Loth}, the rarefaction-specific terms are
\begin{subequations}
    \begin{align}
        C_\mathrm{D, Kn, Re} &= \frac{24}{Re_\mathrm{p}} \left(1 + 0.15 Re_\mathrm{p}^{0.687}\right) f_\mathrm{Kn}, \\
        f_\mathrm{Kn} &= \frac{1}{1+Kn_\mathrm{p} \left[2.514 + 0.8 \,\exp\mathopen{}\left(-\frac{0.55}{Kn_\mathrm{p}}\right)\right]}, \\
        C_\mathrm{D, f_M, Re} &= \frac{C_\mathrm{D, f_M}}{1+\left(\frac{C_\mathrm{D, f_M}}{1.63}-1\right) \sqrt{\frac{Re_\mathrm{p}}{45}}}, \\
        C_\mathrm{D, f_M} &= \frac{\left(1 + 2s_\mathrm{M}^2\right) \mathrm{erf}\mathopen{}\left(-s_\mathrm{M}^2\right)}{s_\mathrm{M}^3 \sqrt{\pi}} + \frac{\left(4s_\mathrm{M}^4 + 4s_\mathrm{M}^2 - 1\right) \mathrm{erf}\mathopen{}\left(s\right)}{2s_\mathrm{M}^4} + \frac{2}{3s_\mathrm{M}} \sqrt{\frac{\pi T_\mathrm{p}}{T_\mathrm{f}}} , \quad\text{and}\\
        s_\mathrm{M} &\equiv Ma_\mathrm{p} \sqrt{\gamma/2},
    \end{align}
\end{subequations}
where $T_\mathrm{p}$ is the particle temperature. Compression-specific terms in Eq.~\eqref{equ:Loth} are
\begin{subequations}
    \begin{align}
        H_\mathrm{M} &= 1 - \frac{0.258 \,C_\mathrm{M}}{1 + 514 \,G_\mathrm{M}}, \\
        G_\mathrm{M} &= \left\{\begin{array}{ll} 1 - 1.525 \,Ma_\mathrm{p}^4, &\quad Ma_\mathrm{p} < 0.89 \\ 0.0002 + 0.0008 \,\mathrm{tanh}\mathopen{}\left[12.77 \left(Ma_\mathrm{p} - 2.02\right) \right], &\quad Ma_\mathrm{p} > 0.89\end{array}\right., \quad\text{and} \\
        C_\mathrm{M} &= \left\{\begin{array}{ll} \frac{5}{3} + \frac{2}{3} \,\mathrm{tanh}\mathopen{}\left[3 \,\log\mathopen{}\left(Ma_\mathrm{p} - 0.1\right) \right], &\quad Ma_\mathrm{p} < 1.45 \\ 2.044 + 0.2 \,\exp\mathopen{}\left[-1.8 \,\log\mathopen{}\left(\frac{Ma_\mathrm{p}}{1.5} \right)^2 \right], &\quad Ma_\mathrm{p} > 1.45\end{array}\right..
    \end{align}
\end{subequations}
The resultant drag law has been extensively benchmarked using experimental data and employed for many simulations of high-speed particle-laden flow. Recently, Loth et al. \cite{Loth2021} published a comprehensive review of relevant results, obtained from particle-resolved DNSs, rarefied-gas simulations, and wind tunnel experiments. The authors found that Loth's original model was not empirically supported near $Re_\mathrm{p} = 45$. They updated the model from \cite{Loth2008} to correct for these discrepancies \cite{Loth2021}. However, updates in the latter paper do not meaningfully affect the cone--cylinder simulation in Sec.~\ref{sec:demos:cone} because, again, it has a maximum $Re_\mathrm{p}$ of about 16. We thus employ the original formulation of Loth, as presented above.\par

\section{Computational minutiae}
\label{app:implementation}
We implement all the neural networks used for this work in TensorFlow~2.10. Prior to training, the weights are drawn from a standard normal distribution and the biases are set to zero. Loss term weighting parameters, $\chi_1$, $\chi_2$, and $\chi_3$, are selected through a simple parameter sweep in a representative synthetic scenario. Training is performed with the Adam optimizer; the initial learning rate of $10^{-3}$ is dropped to $10^{-4}$ after a plateau in $\mathcal{L}_\mathrm{total}$. Mini-batch training with a batch size of 5000 samples is conducted for both the flow and particle physics loss components. Training persists to convergence in all cases, which takes around 30,000 epochs at each learning rate.\par

The networks employed to represent the isotropic turbulent flow have a depth of 15 layers and width of 300 neurons; the networks used for the cone--cylinder flow have a depth of eight layers and width of 150 neurons. Both architectures are empirically chosen to ensure adequate expressivity for their respective flow case. When creating a Fourier layer, $\mathcal{G}$, we draw the frequencies in $\boldsymbol\upomega$ from a centered Gaussian distribution, having a standard deviation of unity for the spatial coordinates in $\mathbf{z}^0$ and 0.2 for time; $w$ is set to 256 throughout this work to promote a broad range of spectral content in $\mathcal{F}$. Based on the above architecture and training protocol, the average reconstruction takes around 20~hours for the isotropic turbulent flow case and 5~hours for the cone--cylinder case, using an NVIDIA\textsuperscript{\textregistered} RTX\textsuperscript{\texttrademark} 3090 GPU with 24GB of memory.\par

Four key numerical tricks are employed for stability. First, we non-dimensionalize all equations in the physics losses, which ensures that the loss components are roughly $\mathcal{O}(1)$, thereby balancing the contribution of each equation to the local step direction. Second, for the cone--cylinder flow model, we formulate a non-singular variant of Eq.~\eqref{equ:cone equations}. All radial partial derivatives of $r\phi$, where $\phi$ is any quantity or product of quantities, are expanded in the usual way,
\begin{equation}
    \frac{\partial(r\phi)}{\partial r} = r\frac{\partial\phi}{\partial r} + \phi. \nonumber
\end{equation}
We then multiply the continuity, axial momentum, and energy equations by $r$ and the radial momentum equation by $r^2$. This formulation tempers the effects of spurious jumps in partial derivatives close to the axis of symmetry (see \cite{Molnar2023b}). Third, we specify a boundary loss at the inlet, using the known free-stream conditions, as well as a slip-wall loss along the cone--cylinder body. These losses prevent reconstruction errors caused by weak solutions to Eq.~\eqref{equ:cone equations}. Fourth, we enforce hard positivity constraints on several physical variables, including $d_\mathrm{p}$, $\rho_\mathrm{f}$, $E_\mathrm{f}$, and $T_\mathrm{f}$. The constraints are implemented by parameterizing these quantities with a positive function, e.g., sigmoid, Softplus, or similar.\par

\section*{Acknowledgements}
This material is based upon work supported by the Erlangen Graduate School in Advanced Optical Technologies at the Friedrich-Alexander-Universit{\"a}t Erlangen-N{\"u}rnberg. The authors thank M. Bross, D. Fries, and T. A. McManus for their feedback.\par

\bibliographystyle{osajnl2} 

\end{document}